\documentclass[11pt]{article}
\usepackage{amssymb}
\usepackage[space]{cite}
\pdfoutput=1
\usepackage{multirow}
\usepackage{hyperref}
\usepackage[pdftex]{color,graphicx}
\renewcommand{\d}{\mathrm{d}}

\begin{document}

\author{A.~Bernotas\,$^{a}$ and V.~\v{S}imonis\,$^{b}$ \and {\small ${}^{a}$~Vilnius University Faculty of Physics,} \and {\small Saul\.{e}tekio 9, LT-10222 Vilnius, Lithuania} \and {\small ${}^{b}$~Vilnius University Institute of Theoretical Physics and Astronomy,} \and {\small A. Go\v{s}tauto 12, LT-01108 Vilnius, Lithuania}}
\title{EFFECT OF THE CASIMIR ENERGY ON THE BAG MODEL PHENOMENOLOGY OF LIGHT HADRONS}
\date{{\small Received {\today}}}
\maketitle

\begin{abstract}
The dependence of light hadron masses and baryon magnetic moments on the magnitude of the Casimir energy is examined in the bag model with center-of-mass corrections. There are seven free parameters in the model. Six of them are determined from the fit to the masses of certain hadrons, and the last one (Casimir energy parameter) from the best fit to the magnetic moments of light baryons. The predicted magnetic moments are compared with the results obtained in various other models and with experiment data.

\medskip


\end{abstract}

\section{Introduction}

Magnetic moments, beside the masses, are fundamental parameters of hadrons and carry significant information about their internal structure. Therefore they have been of theoretical interest for a long time up to now. Various approaches and models can be used to calculate these quantities. However, as noted in \cite{LT11}, the mass spectrum and magnetic moments probe largely orthogonal physical effects. For example, if we have a model adjusted to provide sufficiently good description of hadron masses, it is not certain that the description of magnetic moments will be of the same quality. In QCD we expect that a more accurate solution gives better description of masses and magnetic moments simultaneously. However, the phenomenological models used in practice to calculate hadron properties (various potential, chiral, bag models, etc) are not QCD. The typical example is the MIT bag model. Magnetic moments predicted in Ref.~\cite{DJJK75} are about 30\% too small, in a serious conflict with the experiment. Nevertheless, if one takes the ratios of all the other moments to that of a proton, the bag model predictions are similar to the usual quark model predictions. Thus, the lower absolute value of magnetic moments in the bag model seems to be the overall scale problem. Magnetic moments of individual quarks in the bag are associated with overlaps of the small and large components of the Dirac wave functions. This overlap is proportional to the bag radius $R$. Can we find a way to enlarge the bag radius and not to spoil the relatively good description of the mass spectra? In this case one is naturally tempted to add an extra phenomenological term which may help to ensure larger bag radii and consequently better values of magnetic moments. For example, the alternative confinement scheme based on a surface tension rather than a volume pressure was proposed \cite{HK78}. Can one produce proper values of the magnetic moments by adding the surface energy term $4\pi TR^{2}$ to the volume energy $\frac{4\pi}{3}BR^{3}$ and refitting the model parameters? Unfortunately, no. The model with the surface tension was examined in detail in Ref.~\cite{MK01}. It appears that the model in which the volume energy term is replaced by the surface tension energy provides results very similar to the former one. Thus, if one adds the two terms together, one can arbitrarily choose one of the free parameters ($B$ or $T$), and the results obtained are almost insensitive to this choice. One possible solution to the problem was proposed in \cite{HDDT78}. The point is to add the term that differs between mesons and baryons. The simplest such term is $C\left| N_{q}-N_{\overline{q}}\right|$, where $N_{q}$ and $N_{\overline{q}}$ are the numbers of quarks and antiquarks, and $C$ is a new free parameter which could be adjusted so as to produce the needed large bag radius and ensure the correct value of the magnetic moment of (say) the proton. This procedure improves the overall description of baryon magnetic moments, however, another attractive feature of the bag model~-- the unified treatment of mesons and baryons using the same expression for the particle mass~-- is being lost. A different and more elegant (in our opinion) way to deal with the problem is associated with the so called center-of-mass motion (c.\,m.\,m.) corrections. The bag model is usually constructed as an independent particle shell model. Therefore there is a sizeable spurious contribution to the energy from the motion of the center of mass, which is confined inside the bag. For the ground-state hadrons, the c.\,m.\,m. energy is simply an inconvenience and requires some correction to the energy. While there are questions as to the best method for estimating these corrections, they must be taken into account in some fashion and, as noted in Ref.~\cite{HK82}, may be applied to restore the correct order of magnitude of the magnetic moments. The idea really works. The corrected values of the magnetic moments are much closer to the experimental data \cite{BSRMT84,S98,BS04}. Still, some discrepancy remains indicating that the correction is somewhat too weak. Can we proceed with the improvement of the magnetic moments? In the present paper we will show that such improvement is possible via choosing a proper value of the zero-point (Casimir) energy.

The concept of Casimir energy appears in the bag model phenomenology almost immediately after the birth of the model \cite{DJJK75} with the wrong sign of the energy term at first. In the early versions of the model the bag energy necessary to fit hadron masses had the term $Z_{0}/R$ with negative $Z_{0}$. This term was plausibly erroneously interpreted as the Casimir energy. When the c.\,m.\,m. corrections are taken into account, from the phenomenological point of view there is no need of such a term at all \cite {CHP83}. Moreover, the careful analysis shows \cite{M83} that Casimir force for a spherical shell with the bag model boundary conditions is repulsive, and, consequently, the constant $Z_{0}$ must be of positive sign. Now it is almost clear how Casimir energy may improve the magnitude of magnetic moments. The strength of the c.\,m.\,m. correction is usually derived from the fit of certain hadron masses to the experimental data \cite{S98,BS04}. So, if we add some positive term to the uncorrected energy, the correction would be stronger, enlarging the magnitude of the corrected magnetic moments in turn.

The paper is organized as follows: in Sec.~2 we describe briefly the model we are dealing with and present explicit expressions for magnetic moments of the baryon octet and decuplet. In Sec.~3 we examine the dependence of the light hadron masses and magnetic moments on the magnitude of the Casimir energy parameter $Z_{0}$. The ``best'' fit to the magnetic moments is found. The new model parameters are used to recalculate the ground-state hadron masses and magnetic moments of the light baryon octet and decuplet. The results obtained are compared with other theoretical predictions and experimental data. Discussion and concluding remarks are given in Sec.~4.

\section{MIT bag model in the static spherical cavity approximation}

The MIT bag model was at first formulated as a Lorentz-invariant field theory \cite{CJJTW74}. However, for the investigation of hadron properties the static spherical cavity approximation of the model was widely used and even became the synonym of the MIT bag model. The hadron bag energy in this approximation is given by
\begin{equation}
E=E_{V}+E_{q}+\Delta E\,+E_{0} \, .
\label{eq2.01A}
\end{equation}
The four terms on the right-hand side are:
\begin{itemize}
\item{}
the bag volume energy, 
\begin{equation}
E_{V}=\frac{4\pi }{3}BR^{3} \, ,
\label{eq2.01B}
\end{equation}
where $B$ is the so-called bag constant, and $R$ is the radius of the confinement region (bag radius);
\item{}
the sum of single-particle quark energies
\begin{equation}
E_{q}=\sum\limits_{i}\varepsilon _{i} \, ,
\label{eq2.01C}
\end{equation}
where the energies of individual quarks obey the eigenvalue equation 
\begin{equation}
\tan \left( R\sqrt{\varepsilon^{2}-m^{2}}\right) =\frac{R\sqrt{\varepsilon^{2}-m^{2}}}{1-mR-\varepsilon R} \, ;
\label{eq2.01D}
\end{equation}
\item{}
quark--quark interaction energy due one-gluon exchange 
\begin{equation}
\Delta E=E^{\mathrm{m}}+E^{\mathrm{e}} \, ,
\label{eq2.01E}
\end{equation}
where $E^{\mathrm{m}}$ and $E^{\mathrm{e}}$ are the color-magnetostatic and color-electrostatic (Coulomb) pieces of the interaction energy (for more details see \cite{BS04});
\item{}
Casimir energy term 
\begin{equation}
E_{0}=\frac{Z_{0}}{R} \, .
\label{eq2.01F}
\end{equation}
\end{itemize}

The parameter $Z_{0}$ is thought to be calculable in QCD, and to some extent it is.

The interaction energy in Eq.~(\ref{eq2.01E}) is computed to the first order in the scale-dependent effective strong coupling constant 
\begin{equation}
\alpha _{\mathrm{c}}(R)=\frac{2\pi }{9\ln (A+R_{0}/R)} \, ,
\label{eq2.02}
\end{equation}
where $R_{0}$ is the scale parameter which plays the role similar to QCD constant ($R_{0}\sim 1/\Lambda $), and parameter $A$ serves to avoid divergences in the case $R\rightarrow R_{0}$. Up and down quarks are assumed to be massless, and the (scale-dependent) mass of the strange quark is to be obtained from the mass function 
\begin{equation}
\overline{m}_{s}(R)=\widetilde{m}_{s}+\alpha _{\mathrm{c}}(R)\cdot \delta_{s} \, , 
\label{eq2.03}
\end{equation}
where $\widetilde{m}_{s}$ and $\delta _{s}$ are two free model parameters.

The calculation of the hadron mass spectrum is performed in two steps. First, for each hadron the energy (\ref{eq2.01A}) is to be minimized with respect to the bag radius $R$. In this way the bag radii $R_{i}$ of individual hadrons are obtained. Then one can use Eq.~(\ref{eq2.01A}) to calculate the hadron bag energy $E_{i}$. However, this is not the whole story. The bag energy still contains the spurious c.\,m.\,m. energy. A prescription must be given to relate the eigenvalues of the static bag model Hamiltonian to the masses of hadrons. To this end we adopt the procedure proposed in Ref.~\cite{DJ80} and applied in \cite{BSMT84,TT85,HM90}. In this approach the bag state is expressed as the wave packet of the physical states $\left| H, \mathbf{k}\right\rangle $ with various total momenta: 
\begin{equation}
\left| B\right\rangle =\int {\d}^{3}k \, \Phi _{P}(|\mathbf{k}|)\left| H, \mathbf{k}\right\rangle \, ,
\label{eq2.04A}
\end{equation}
with the Gaussian parametrization of the profile function \cite{BS04,HM90} 
\begin{equation}
\Phi _{P}(s)=\left( \frac{3}{2\pi P^{2}}\right)^{3/4}\,\exp \left( -\frac{3s^{2}}{4P^{2}}\right) \, .
\label{eq2.04B}
\end{equation}

The effective momentum square $P^{2}$ is defined as
\begin{equation}
P^{2}=\gamma \sum\limits_{i}p_{i}^{2} \, ,
\label{eq2.04C}
\end{equation}
where $p_{i}=\left( \varepsilon _{i}^{2}-m_{i}^{2}\right)^{1/2}$ are the momenta of the individual quarks, and $\gamma$ is an adjustable parameter governing the c.\,m.\,m. correction.

The relation between the bag model energy $E$ and the mass $M$ of a particular hadron is given by 
\begin{equation}
E=\int {\d}^{3}s\,\Phi _{P}^{2}(s)\,\sqrt{M^{2}+s^{2}} \, .
\label{eq2.04D}
\end{equation}

In order to obtain the mass of the particle, Eq.~(\ref{eq2.04D}) is to be solved numerically. We also must decide how to deal with the zoo of free parameters. Altogether,  there are seven parameters~-- $B$, $\gamma $, $A$, $R_{0}$, $\widetilde{m}_{s}$, $\delta _{s}$, and $Z_{0}$. We will use the same prescription as in \cite{BS04} for the first six of them: $B$, $\gamma $, $A$, and $R_{0}$ will be determined by fitting calculated masses of light hadrons ($N$, $\Delta $, $\pi $, and the average mass of the $\omega$--$\rho$ system) to experimental data, $\widetilde{m}_{s}$ and $\delta_{s}$ from the fit to the masses of $\phi $ and $\Lambda $. There remains one more parameter $Z_{0}$, which scales the magnitude of the Casimir energy. All questions associated with this term we postpone for the next section.

The last ingredient we need for our investigation is the expressions for the baryon magnetic moments. In the bag model, just as in the simple quark model, magnetic moments of baryons can be represented as 
\begin{equation}
\mu_{\mathrm{H}}^{0}=\sum\limits_{i}\mu_{i}\left\langle \mathrm{H\uparrow }\right| \mathbf{\sigma }_{z}^{i}\left| \mathrm{H\uparrow }\right\rangle \, .
\label{eq2.05A}
\end{equation}
The magnetic moments of quarks confined in the bag have the form 
\begin{equation}
\mu_{i}=q_{i}\ \bar{\mu}_{i} \, ,
\label{eq2.05B}
\end{equation}
where $q_{i}$ is the quark electric charge, and reduced (charge independent) quark magnetic moments $\bar{\mu}_{i}$ are given by \cite{DJJK75} 
\begin{equation}
\bar{\mu}_{i}=\frac{4\varepsilon _{i}R_{\mathrm{H}}+2m_{i}R_{\mathrm{H}}-3}{2(\varepsilon _{i}R_{\mathrm{H}}-1)\varepsilon _{i}R_{\mathrm{H}}+m_{i}R_{\mathrm{H}}}\ \frac{R_{\mathrm{H}}}{6} \, .
\label{eq2.05C}
\end{equation}

Magnetic transition moments are defined by 
\begin{equation}
\mu_{\mathrm{H\rightarrow H}^{\prime }}^{0}=\sum\limits_{i}\mu_{i}\big\langle \mathrm{H}^{\prime }\mathrm{\uparrow }\big\vert\, \mathbf{\sigma }_{z}^{i}\,\big\vert \mathrm{H\uparrow}\big\rangle \, ,
\label{eq2.05D}
\end{equation}
where for simplicity $R_{\mathrm{H}}=R_{\mathrm{H}^{\prime }}$ is assumed.

\begin{table}[tbp]
\centering
\caption{Composition of baryon octet magnetic moments in terms of magnetic moments of individual quarks (column~2) and in terms of corresponding reduced quantities (column~3).
\label{t2.1}} 
\begin{tabular}{lcc}
\hline
Particles & $\mu_{\mathrm{H}}^{0}$ & $\mu_{\mathrm{H}}^{0}$ \\
\hline \\[-9pt]
$P$ & $\frac{1}{3}(4u-d)$ & $\bar{u}$ \\[2ex]
$N$ & $\frac{1}{3}(4d-u)$ & $-\frac{2}{3}\bar{u}$ \\[2ex]
$\Lambda $ & $s$ & $-\frac{1}{3}\bar{s}$ \\[2ex]
$\Sigma^{0}$ & $\frac{1}{3}(2u+2d-s)$ & $\frac{1}{9}(2\bar{u}+\bar{s})$ \\[2ex]
$\Sigma^{-}$ & $\frac{1}{3}(4d-s)$ & $\frac{1}{9}(\bar{s}-4\bar{u})$ \\[2ex]
$\Sigma^{+}$ & $\frac{1}{3}(4u-s)$ & $\frac{1}{9}(8\bar{u}+\bar{s})$ \\[2ex]
$\Xi^{-}$ & $\frac{1}{3}(4s-d)$ & $\frac{1}{9}(\bar{u}-4\bar{s})$ \\[2ex]
$\Xi^{0}$ & $\frac{1}{3}(4s-u)$ & $-\frac{2}{9}(\bar{u}+2\bar{s})$ \\[2ex]
$\Sigma \rightarrow \Lambda $ & $\frac{1}{\sqrt{3}}(d-u)$ & $-\frac{1}{\sqrt{3}}\bar{u}$ \\[1ex]
\hline
\end{tabular}
\end{table}

\begin{table}[tbp]
\centering
\caption{Composition of baryon decuplet magnetic moments in terms of magnetic moments of individual quarks (column~2) and in terms of corresponding reduced quantities (column~3).\label{t2.2}} 
\begin{tabular}{lcc}
\hline
Particles & $\mu_{\mathrm{H}}^{0}$ & $\mu_{\mathrm{H}}^{0}$ \\
\hline \\[-9pt]
$\Delta^{-}$ & $3d$ & $-\bar{u}$ \\[2ex]
$\Delta^{0}$ & $2d+u$ & $0$ \\[2ex]
$\Delta^{+}$ & $2u+d$ & $\bar{u}$ \\[2ex]
$\Delta^{++}$ & $3u$ & $2\bar{u}$ \\[2ex]
$\Sigma^{*\ -}$ & $2d+s$ & $-\frac{1}{3}(2\bar{u}+\bar{s})$ \\[2ex]
$\Sigma^{*\ 0}$ & $u+d+s$ & $\frac{1}{3}(\bar{u}-\bar{s})$ \\[2ex]
$\Sigma^{*\ +}$ & $2u+s$ & $\frac{1}{3}(4\bar{u}-\bar{s})$ \\[2ex]
$\Xi^{*\ -}$ & $2s+d$ & $-\frac{1}{3}(\bar{u}+2\bar{s})$ \\[2ex]
$\Xi^{*\ 0}$ & $2s+u$ & $\frac{2}{3}(\bar{u}-\bar{s})$ \\[2ex]
$\Omega^{-}$ & $3s$ & $-\bar{s}$ \\
\hline
\end{tabular}
\end{table}

The origin of magnetic moments in the bag model is a rather interesting phenomenon by itself. Massless structureless Dirac particles (\textit{u-} and \textit{d-}quarks) have no intrinsic magnetic moments at all. Therefore the magnetic moments of light baryons (proton, neutron, etc) are as they are only because of the confinement. But should we wonder? The proton consisting of massless quarks has a nonzero mass only because of the confinement as well.

Matrix elements $\left\langle \mathrm{H}^{\prime }\mathrm{\uparrow }\right| \mathbf{\sigma }_{z}^{i}\left| \mathrm{H\uparrow }\right\rangle $ can be readily calculated with SU(6) wave functions providing the usual quark model expressions for the baryon magnetic moments. The results are presented in Tables~\ref{t2.1} and \ref{t2.2} for the baryon octet ($J=1/2$) and decuplet ($J=3/2$), respectively. For simplicity, in cases with no ambiguity the shorthand notations ($\mu_{P}\rightarrow P$, $\bar{\mu}_{s}\rightarrow \bar{s}$, etc) are used. The entries in columns~3 were obtained assuming isospin symmetry, i.\,e. $\bar{\mu}_{u}=\bar{\mu}_{d}$ (or $\bar{u}=\bar{d}$ in the shorthand notations).

From Tables~\ref{t2.1} and \ref{t2.2} several quark model relations can be deduced immediately:
\begin{eqnarray}
N &=&-\frac{2}{3}P \, ,
\label{eq2.06A1} \\[2ex]
\Sigma^{0} &=&\phantom{-}\frac{1}{2}(\Sigma^{+}+\Sigma^{-}) \, ,
\label{eq2.06A2} \\[2ex]
\Delta^{-} &=&-\Delta^{+} \, ,
\label{eq2.06A3} \\[2ex]
\Delta^{++} &=&\phantom{-}2\Delta^{+} \, ;
\label{eq2.06A4} \\[2ex]
\Omega^{-} &\simeq &\phantom{-}3\Lambda \, ,
\label{eq2.06B1} \\[2ex]
\Xi^{*\ 0} &\simeq &\phantom{-}2\Sigma^{*\ 0} \, ,
\label{eq2.06B2} \\[2ex]
\Sigma^{*\ -} &\simeq &-3\Sigma^{0} \, .
\label{eq2.06B3}
\end{eqnarray}

In the bag model (without corrections) the relations (\ref{eq2.06A1})--(\ref{eq2.06A4}) hold exactly, while (\ref{eq2.06B1})--(\ref{eq2.06B3}) are only approximate because the magnetic moments of quarks depend on the bag radius which for particles entering Eqs.~(\ref{eq2.06B1})--(\ref{eq2.06B3}) differ ($R_{\Omega^{-}}\neq R_{\Lambda }$, etc). The famous quark model relation $\mu_{N}/\mu_{P} =$ $-$2$/$3 (Eq.~(\ref{eq2.06A1})) differs from the experimental value $-$0.68 by about 3\%, a typical discrepancy that could be caused by the isospin symmetry breaking.

Before comparing the quantities computed in the static spherical cavity approximation with experimental data, they must be corrected for the center-of-mass motion. Maybe the simplest (though, plausibly not very accurate) way to do this is to adopt the prescription proposed by A.~Halprin and A.K.~Kerman \cite{HK82}. To make things as clear as possible we repeat their derivation below. Let us assume that the c.\,m.\,m. corrected bag energy $E_{\mathrm{cor}}$ (to be identified with the mass of the particle $E_{\mathrm{cor}}\rightarrow M$) is given by the relation 
\begin{equation}
E_{\mathrm{cor}}^{2}=E_{\mathrm{bag}}^{2}-P^{2} \, ,
\label{eq2.07}
\end{equation}
where $E_{\mathrm{bag}}=E$ is the static cavity bag energy, and $P^{2}=\left\langle P^{2}\right\rangle$ is the expectation of the mean-squared total momentum of the system given by Eq.~(\ref{eq2.04C}). In the presence of the magnetic field, if $\left\langle P^{2}\right\rangle$, at least to the first order, is unaffected, we have
\begin{equation}
E_{\mathrm{cor}}^{2}(H)=E_{\mathrm{bag}}^{2}(H)-P^{2} \, .
\label{eq2.08}
\end{equation}

Differentiation of both sides of the last equation with respect to $H$ and evaluation in the weak-field limit yields 
\begin{equation}
2E_{\mathrm{cor}}\frac{\partial E_{\mathrm{cor}}(H)}{\partial H}\thickapprox 2E_{\mathrm{bag}}\frac{\partial E_{\mathrm{bag}}(H)}{\partial H} \, .
\label{eq2.09}
\end{equation}

Now, because $\frac{\partial E_{\mathrm{cor}}(H)}{\partial H}$ and $\frac{\partial E_{\mathrm{bag}}(H)}{\partial H}$ in the limit $H\rightarrow 0$ are, respectively, corrected ($\mu $) and uncorrected ($\mu^{0}$) magnetic moments, we arrive at the expression 
\begin{equation}
\mu =\frac{E}{M}\, \mu^{0} \, ,
\label{eq2.10}
\end{equation}
where we have returned to our previous notations $E_{\mathrm{cor}}\rightarrow M$, $E_{\mathrm{bag}}=E$. It is plausible that by applying Eq.~(\ref{eq2.10}) the c.\,m.\,m. corrections could be to some extent overestimated. On the other hand, this relation is very attractive because of its simplicity and universality. Note, however, that Eq.~(\ref{eq2.07}) used in the derivation of the relation~(\ref{eq2.10}) differs from our previous choice (Eq.~(\ref{eq2.04D})). For further applications we need to know to what extent Eq.~(\ref{eq2.10}) is compatible with the c.\,m.\,m. corrections defined via Eq.~(\ref{eq2.04D}). As noted in \cite{HM90}, Eq.~(\ref{eq2.04D}) may be rewritten in the form 
\begin{equation}
M^{2}=E^{2}-\beta \left( \frac{M^{2}}{P^{2}}\right) P^{2} \, ,
\label{eq2.11A}
\end{equation}
where 
\begin{equation}
\beta (x)=\frac{54}{\pi }\bigg[ \int\limits_{0}^{\infty }t^{2} \d t \, \sqrt{t^{2}+x}\,\exp \bigg( -\frac{3}{2}t^{2}\bigg) \bigg]^{2}-x \, .
\label{eq2.11B}
\end{equation}

In the nonrelativistic case the function $\beta (x)$ approaches 1, and Eq.~(\ref{eq2.04D}) (or Eq.~(\ref{eq2.11A})) is equivalent to the relation 
\begin{equation}
M^{2}=E^{2}-P^{2} \, ,
\label{eq2.12}
\end{equation}
which is nothing else than Eq.~(\ref{eq2.07}). In the case of light hadrons we are dealing with values of $\beta (x)$ lying in the range 0.925--0.945 (see Ref.~\cite{BS04}), though the use of Eq.~(\ref{eq2.12}) instead of Eq.~(\ref{eq2.04D}) may introduce an error of about 10\%. So, strictly speaking, if one is going to calculate c.\,m.\,m. corrected magnetic moments via Eq.~(\ref{eq2.10}), the better prescription for the hadron mass should be Eq.~(\ref{eq2.12}). On the other hand, Eq.~(\ref{eq2.04D}) seems to be preferable from the theoretical point of view. In the end, both them are just prescriptions and after refitting the model parameters could give to some extent similar results. Nevertheless, the approach based on Eqs.~(\ref{eq2.04A}) and (\ref{eq2.04D}) has one additional advantage because in this case for spin-1$/$2 baryons one has more refined formula for the c.\,m.\,m. corrected magnetic moments \cite{BSMT84} 
\begin{equation}
\mu_{\mathrm{\mathrm{cor}}}=\frac{3}{1+\left\langle M/E\right\rangle +\left\langle M^{2}/E^{2}\right\rangle }\bigg( \mu^{0}+\frac{1-\left\langle M/E\right\rangle }{3}\frac{M_{\mathrm{P}}}{M}Q\bigg) \, ,
\label{eq2.13}
\end{equation}
where $\mu^{0}$ is uncorrected magnetic moment, $M_{\mathrm{P}}$ is mass of the proton, $M$ and $Q$ is mass and charge of the baryon under consideration. The averages $\left\langle M/E\right\rangle$ and $\left\langle M^{2}/E^{2}\right\rangle$ have to be calculated with the profile $\Phi_{P}(s)$ given by Eq.~(\ref{eq2.04B}). Since the relation (\ref{eq2.13}) was derived for the specific case of $S=1/2$ baryons, it cannot be used directly in the case of $S =$ 3$/$2. The analysis of the spin-3$/$2 fermions implies the use of the Rarita--Schwinger spinors, which causes additional complications. However, we still have a more universal relation (\ref{eq2.10}) at our disposal.

\section{Reintroducing Casimir energy}

It is almost a common agreement that the bag model, if taken seriously, must contain a Casimir energy term. If one naively tried to generalize the QED result, one would readily obtain $Z_{0}\thickapprox$ 0.37 (see Ref.~\cite{M01} for discussion). However, such calculation includes contributions from both exterior and interior gluon field modes, but only the latter~-- because of the confinement~-- should be considered in the bag model. The correct bag model result found by the Green function method \cite{M83} has logarithmic divergence 
\begin{equation}
E_{0}=\frac{8}{R}\bigg[0.090+0.0081\ln \frac{\theta }{8}\bigg] \, ,
\label{eq3.1}
\end{equation}
where $\theta \rightarrow$ 0 is a cutoff parameter which could be associated with the bag ``skin depth'' representing a realistic boundary, instead of a sharp mathematical one. As suggested by K.A.~Milton \cite{M83}, Eq.~(\ref {eq3.1}) may be used in bag model calculations with an effective $\theta \ll$~1.

A very similar result was obtained in \cite{LR96} using zeta function method for the regularized energy mode summation,
\begin{equation}
E_{0}=\frac{8}{R}\,[0.084 + 0.0081 \ln(\xi R)] \, .
\label{eq3.2}
\end{equation}

This expression contains the energy scale parameter $\xi $ which in pure QCD should be associated with the QCD constant $\Lambda $.

We see that in any case the Casimir energy term can be expressed as $Z_{0}/R$ with the model dependent parameter $Z_{0}$, the values of which vary in the interval 0~$\leqslant Z_{0}\leqslant$~1. In previous bag model calculations \cite{BS04} we have ignored the Casimir energy contribution. One reason for this was very simple~-- we wanted to reduce the number of free model parameters. The careful reader could find a more serious objection. For example, because the Casimir energy term in the bag model Hamiltonian acts as certain stabilizing factor, in the presence of Casimir energy the empty bags (without quarks and gluons) are allowed. Could one imagine such lumps of energy travelling across the universe? Our opinion is that this must not be a very severe problem. We already know that the bag model can contain spurious states (e.\,g., orbital excitations of the center-of-mass). The origin of the empty bag state seems to be the same as the origin of the Casimir energy~-- vacuum fluctuations. Evidently, it is in the spirit of the bag model philosophy that even the vacuum fluctuations of the gluon field are confined in the bag and therefore such solutions are almost unavoidable. The vacuum fluctuations are not real physical states, so we think that an empty bag state can be interpreted as spurious and safely ignored. In a sense it is the manifestation of the nontrivial structure of QCD vacuum, pointing out explicitly that the perturbative vacuum of the bag model is not the ground state of the true physical vacuum.

Since the presence of Casimir energy in the bag model seems to have rather firm theoretical ground, let us see if the reintroduction of this term can improve the model predictions. For a series of values of the Casimir energy parameter $Z_{0}$ in the range 0--1 we calculated the spectrum of light hadrons and baryon magnetic moments. For every value of $Z_{0}$ the model parameters were refitted along the procedure discussed in the previous section and completely analogous to the one applied in \cite{BS04}. In order to get some feeling to what extent the results are model dependent, we performed our analysis in two slightly different variants of the model: one (\textit{Var1}) in which the hadron mass was related to the bag energy via Eq.~(\ref{eq2.12}), and other (\textit{Var2}) in which such relation had the form given by Eq.~(\ref{eq2.04D}). In each case we tried to find the ``best'' value of $Z_{0}$. In order to compare different fits, we used the root mean square deviations between predicted and experimental values of physical
quantities: 
\begin{equation}
\chi (E)=\bigg[ \frac{1}{14}\sum\limits_{i=1}^{14}\big( M_{i}-M_{i}^{\mathrm{ex}}\big)^{2}\bigg]^{1/2} \,
\label{eq3.3}
\end{equation}
for hadron masses and 
\begin{equation}
\chi (\mu )=\bigg[ \frac{1}{9}\sum\limits_{i=1}^{9}\big( \mu_{i}-\mu_{i}^{\mathrm{ex}}\big)^{2}\bigg]^{1/2} \,
\label{eq3.4}
\end{equation}
for magnetic moments. In Eq.~(\ref{eq3.3}) the mass values of 14 hadrons (all light ground state hadrons except $\eta $ and $\eta^{\prime }$ mesons, masses of which cannot be predicted in the lowest order approximation) were used. The summation in Eq.~(\ref{eq3.4}) includes magnetic moments of seven spin-1$/$2 baryons (namely, $P$, $N$, $\Lambda$, $\Sigma^{+}$, $\Sigma^{-}$, $\Xi^{0}$, and $\Xi^{-}$), the $\Sigma^{0}\rightarrow \Lambda $ transition moment, and magnetic moment of $\Omega^{-}$ ~-- the only spin-3$/$2 baryon, the magnetic moment of which has been measured with sufficient precision. In the case of \textit{Var1} we used the Halprin--Kerman relation (Eq.~(\ref{eq2.10})) to calculate c.\,m.\,m. corrected values of magnetic moments. For the \textit{Var2} our choice is more complicated (and possibly not so consistent). In this case for spin-1$/$2 baryons we can use a plausibly more accurate expression (\ref{eq2.13}), and we do. For spin-3$/$2 baryons, in the absence of something better, we make a step aside from the purity requirements and apply the same universal Halprin-Kerman relation as used before.

\begin{table}[tbp]
\centering
\caption{Model parameters in the two variants of the bag model (\textit{Var1} and \textit{Var2})~-- with and without ($Z_{0}=0$) the Casimir energy. Mass parameters ($\widetilde{m},\delta $) are in GeV, $R_{0}$ in GeV$^{-1}$, $B$ in GeV$^{4}$.\label{t3.1}} 
\begin{tabular}{ccccc}
\hline
\multirow{2}{*}{Parameter} & \multicolumn{2}{c}{\textit{Var1}} & \multicolumn{2}{c}{\textit{Var2}} \\
\cline{2-3} \cline{4-5} \\[-12pt] 
 & $Z_{0}=0$ & $Z_{0}=0.22$ & $Z_{0}=0$ & $Z_{0}=0.64$ \\
\hline
$B$$\times$10$^{4}$ & 7.301 & 7.468 & 7.597 & 8.594 \\ 
$\gamma $ & 1.785 & 2.153 & 1.958 & 3.300 \\ 
$A$ & 0.772 & 0.651 & 1.070 & 0.776 \\ 
$R_{0}$ & 3.876 & 4.528 & 2.543 & 4.210 \\ 
$\widetilde{m}_{s}$ & 0.217 & 0.262 & 0.161 & 0.335 \\ 
$\delta_{s}$ & 0.109 & 0.083 & 0.156 & 0.046 \\
\hline
\end{tabular}
\end{table}

\begin{table}[tbp]
\centering
\caption{Masses of light hadrons (in GeV) in the two variants of the bag model (\textit{Var1} and \textit{Var2})~-- with and without ($Z_{0}=0$) the Casimir energy.\label{t3.2}}
\begin{tabular}{cccccc}
\hline
\multirow{2}{*}{Hadrons} & \multicolumn{2}{c}{\textit{Var1}} & \multicolumn{2}{c}{\textit{Var2}} & \multirow{2}{*}{EXP~\cite{PDG12}} \\
\cline{2-3} \cline{4-5} \\[-12pt] 
&$Z_{0}=0$ & $Z_{0}=0.22$ & $Z_{0}=0$ & $Z_{0}=0.64$ & \\
\hline
$\pi $ & 0.137 & 0.137 & 0.137 & 0.137 & 0.137 \\ 
$\rho $ & 0.776 & 0.776 & 0.776 & 0.776 & 0.768 \\ 
$\omega $ & 0.776 & 0.776 & 0.776 & 0.776 & 0.783 \\ 
$N$ & 0.939 & 0.939 & 0.939 & 0.939 & 0.939 \\ 
$\Delta $ & 1.232 & 1.232 & 1.232 & 1.232 & 1.232 \\ 
$K$ & 0.453 & 0.458 & 0.437 & 0.460 & 0.496 \\ 
$K^{*}$ & 0.895 & 0.894 & 0.897 & 0.891 & 0.894 \\ 
$\phi $ & 1.019 & 1.019 & 1.019 & 1.019 & 1.019 \\ 
$\Lambda $ & 1.116 & 1.116 & 1.116 & 1.116 & 1.116 \\ 
$\Sigma $ & 1.158 & 1.159 & 1.159 & 1.159 & 1.193 \\ 
$\Sigma^{*}$ & 1.385 & 1.383 & 1.388 & 1.383 & 1.385 \\ 
$\Xi $ & 1.310 & 1.313 & 1.310 & 1.317 & 1.318 \\ 
$\Xi^{*}$ & 1.537 & 1.536 & 1.543 & 1.539 & 1.533 \\ 
$\Omega^{-}$ & 1.688 & 1.690 & 1.695 & 1.698 & 1.672 \\ 
$\chi (E)$ & 0.016 & 0.015 & 0.020 & 0.015 & --- \\
\hline
\end{tabular}
\end{table}

\begin{table}[tbp]
\centering
\caption{Magnetic moments of light baryons (in nuclear magnetons) in the two variants of the bag model (\textit{Var1} and \textit{Var2})~-- with and without ($Z_{0}=0$) the Casimir energy.\label{t3.3}} 
\begin{tabular}{crrrrr@{}l}
\hline
\multirow{2}{*}{Baryons} & \multicolumn{2}{c}{\textit{Var1}} & \multicolumn{2}{c}{\textit{Var2}} &  & \multirow{2}{*}{EXP~\cite{PDG12}} \\
\cline{2-3} \cline{4-5} \\[-12pt] 
 & \multicolumn{1}{c}{$Z_{0} =$ 0} & \multicolumn{1}{c}{$Z_{0} =$ 0.22} & \multicolumn{1}{c}{$Z_{0} =$ 0} & \multicolumn{1}{c}{$Z_{0} =$ 0.64} & \\
\hline
$P$ & 2.732 & 2.885 & 2.608 & 2.915 & 2. & 793 \\ 
$N$ & $-$1.821 & $-$1.924 & $-$1.658 & $-$1.830 & $-$1. & 913 \\ 
$\Lambda $ & $-$0.598 & $-$0.625 & $-$0.555 & $-$0.607 & $-$0. & 613$\pm$0.004 \\ 
$\Sigma^{+}$ & 2.436 & 2.570 & 2.342 & 2.628 & 2. & 458 \\ 
$\Sigma^{0}$ & 0.756 & 0.796 & 0.701 & 0.776 &  & --- \\ 
$\Sigma^{-}$ & $-$0.924 & $-$0.978 & $-$0.941 & $-$1.075 & $-$1. & 160$\pm$0.025 \\ 
$\Xi^{0}$ & $-$1.307 & $-$1.371 & $-$1.217 & $-$1.346 & $-$1. & 250$\pm$0.014 \\ 
$\Xi^{-}$ & $-$0.496 & $-$0.513 & $-$0.539 & $-$0.601 & $-$0. & 651$\pm$0.003 \\ 
$\left| \Sigma^{0}\rightarrow \Lambda \right| $ & 1.465 & 1.547 & 1.444 & 1.521 & 1. & 61$\pm$0.08 \\ 
$\Omega^{-}$ & $-$1.598 & $-$1.657 & $-$1.587 & $-$1.772 & $-$2. & 02$\pm$0.05 \\ 
$\Delta^{++}$ & 4.846 & 5.040 & 4.829 & 5.389 & 3. & 7\,---\,7.5 \\ 
$\Delta^{+}$ & 2.423 & 2.520 & 2.414 & 2.694 &  & --- \\ 
$\Delta^{0}$ & \multicolumn{1}{r}{0\phantom{.000}} & \multicolumn{1}{r}{0\phantom{.000}} & \multicolumn{1}{r}{0\phantom{.000}} & \multicolumn{1}{r}{0\phantom{.000}} &  & --- \\ 
$\Delta^{-}$ & $-$2.423 & $-$2.520 & $-$2.414 & $-$2.694 &  & --- \\ 
$\Sigma^{*\ +}$ & 2.602 & 2.721 & 2.589 & 2.927 &  & --- \\ 
$\Sigma^{*\ 0}$ & 0.240 & 0.255 & 0.239 & 0.275 &  & --- \\ 
$\Sigma^{*\ -}$ & $-$2.122 & $-$2.211 & $-$2.112 & $-$2.377 &  & --- \\ 
$\Xi^{*\ 0}$ & 0.454 & 0.488 & 0.450 & 0.536 &  & --- \\ 
$\Xi^{*\ -}$ & $-$1.847 & $-$1.923 & $-$1.836 & $-$2.067 &  & --- \\ 
$\chi(\mu)$ & 0.181 & 0.158 & 0.209 & 0.124 &  & --- \\
\hline
\end{tabular}
\end{table}

\begin{figure}
\begin{center}
\includegraphics[scale=1.0]{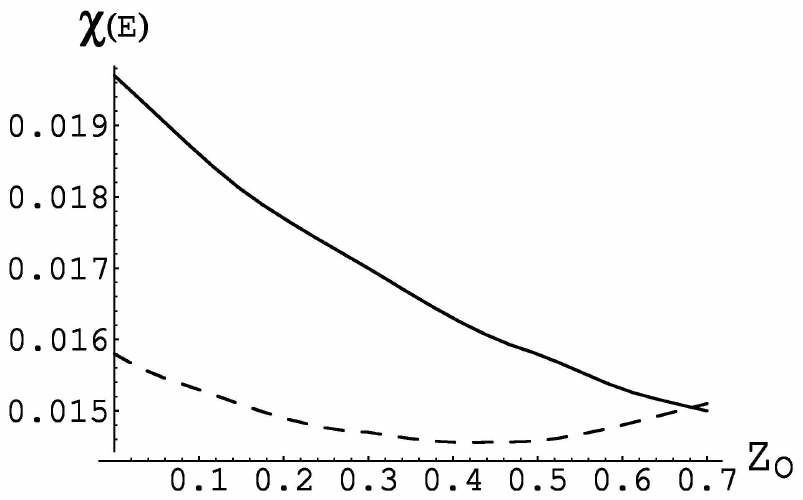}
\end{center}
\caption{Dependence of $\chi(E)$ (in GeV) on $Z_{0}$ for the two variants of model as described in the text. Dashed curve corresponds to \textit{Var1} and the solid one to \textit{Var2}.}
\label{fig3.1}
\end{figure}

\begin{figure}
\begin{center}
\includegraphics[scale=1.0]{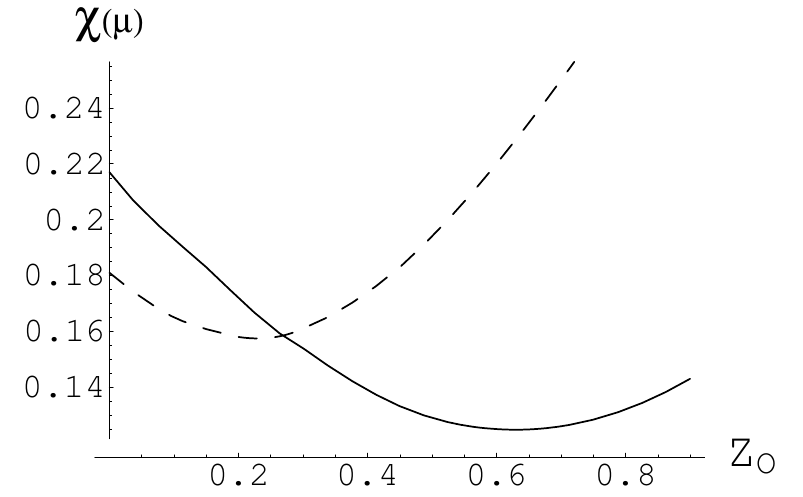}
\end{center}
\caption{Dependence of $\chi(\mu)$ (in nuclear magnetons) on $Z_{0}$ for the two variants of model as described in the text. Dashed curve corresponds to \textit{Var1} and the solid one to \textit{Var2}.}
\label{fig3.2}
\end{figure}

\begin{figure}
\begin{center}
\includegraphics[scale=1.0]{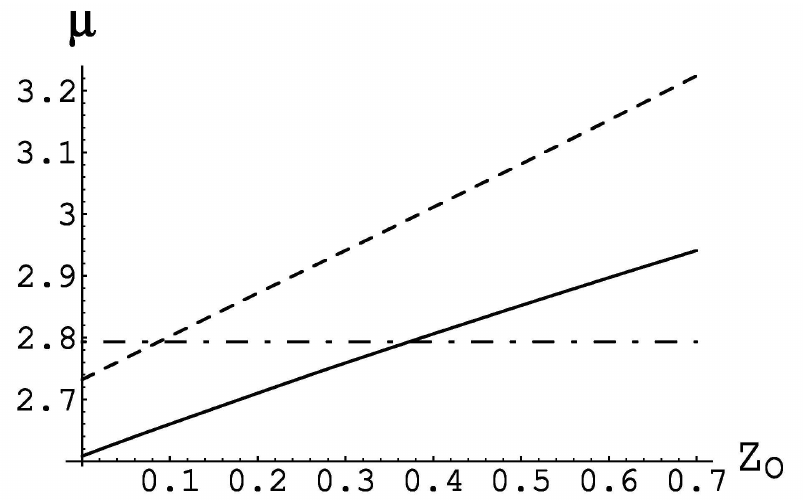}
\end{center}
\caption{Dependence of c.\,m.\,m. corrected magnetic moment of the proton (in nuclear magnetons) on $Z_{0}$ for the two variants of model as described in the text. Dashed curve corresponds to \textit{Var1}, the solid one to \textit{Var2}, and the dash-dotted line denotes the experimental value 2.79.}
\label{fig3.3}
\end{figure}

\begin{figure}
\begin{center}
\includegraphics[scale=1.0]{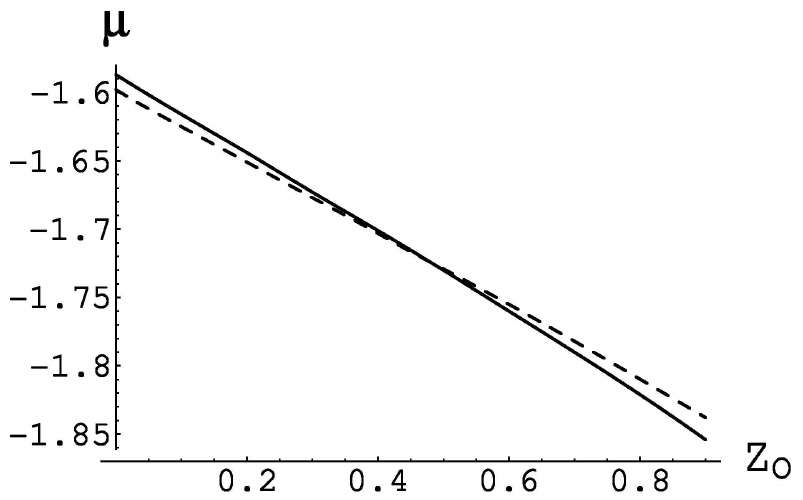}
\end{center}
\caption{
Dependence of c.\,m.\,m. corrected magnetic moment of $\Omega^{-}$ (in nuclear magnetons) on $Z_{0}$ for the two variants of model as described in the text. Dashed curve corresponds to \textit{Var1}, the solid one to \textit{Var2}.}
\label{fig3.4}
\end{figure}

Our results for the masses and magnetic moments are presented in Figs.~\ref{fig3.1}--\ref{fig3.4}, and Tables~\ref{t3.1}--\ref{t3.3}. All experimental data are from Particle Data Tables~\cite{PDG12}. From Figs.~\ref{fig3.1}--\ref{fig3.3} we can make immediate conclusion that, if one does not want to include Casimir energy in the Hamiltonian of the bag model (i.\,e., $Z_{0}=0$), it is preferable to use the first variant of the model (\textit{Var1}) with the c.\,m.\,m. corrections given by the Eqs.~(\ref{eq2.10}) and (\ref{eq2.12}) because it gives better predictions for masses and magnetic moments than the second variant (\textit{Var2} with $Z_{0} =$~0).

Now let us ``switch on'' Casimir force and see what is the effect of the Casimir energy on the masses of hadrons.

\begin{table}[tbp]
\centering
\caption{Magnetic moments (in nuclear magnetons) obtained from the best fit in the bag model (Bag) and in other approaches as described in the text. The quantities used as input data are indicated by asterisk (*).\label{t3.4}}
\begin{tabular}{cr@{}lr@{}lr@{}lr@{}lr@{}lr@{}lr@{}l}
\hline
Particles & \multicolumn{2}{c}{EXP~\cite{PDG12}} & \multicolumn{2}{c}{Bag} & \multicolumn{2}{c}{Nonrel} & \multicolumn{2}{c}{\cite{PRS94,SPN95}} & \multicolumn{2}{c}{\cite{SDCG10}} & \multicolumn{2}{c}{\cite{GMAV08,GMV09}} & \multicolumn{2}{c}{\cite{H07}} \\
\hline
$P$ & 2. & 793 & 2. & 915 & 2. & 724 & 2. & 794 & 2. & 80 & 2. & 58 & 2. & 79* \\ 
$N$ & $-$1. & 913 & $-$1. & 830 & $-$1. & 816 & $-$1. & 894 & $-$2. & 11 & $-$2. & 10 & $-$1. & 91*  \\ 
$\Lambda $ & $-$0. & 613$\pm$0.004 & $-$0. & 607 & $-$0. & 592 & $-$0. & 612 & $-$0. & 58 & $-$0. & 66 & $-$0. & 51 \\ 
$\Sigma^{+}$ & 2. & 458 & 2. & 628 & 2. & 618 & 2. & 68 & 2. & 39 & 2. & 49 & 2. & 46* \\ 
$\Sigma^{0}$ &  & --- & 0. & 776 & 0. & 803 & 0. & 79 & 0. & 54 & 0. & 66 & 0. & 65 \\ 
$\Sigma^{-}$ & $-$1. & 160$\pm$0.025 & $-$1. & 075 & $-$1. & 013 & $-$1. & 088 & $-$1. & 32 & $-$1. & 10 & 1. & 16* \\ 
$\Xi^{0}$ & $-$1. & 250$\pm$0.014 & $-$1. & 346 & $-$1. & 394 & $-$1. & 45 & $-$1. & 24 & $-$1. & 27 & 1. & 25* \\ 
$\Xi^{-}$ & $-$0. & 651$\pm$0.003 & $-$0. & 601 & $-$0. & 487 & $-$0. & 487 & $-$0. & 50 & $-$0. & 95 & $-$1. & 07 \\ 
$\left| \Sigma^{0}\rightarrow \Lambda \right| $ & 1. & 61$\pm$0.08 & 1. & 521 & 1. & 570 & 1. & 6 & 1. & 60 & 1. & 58 &  & --- \\ 
$\Omega^{-}$ & $-$2. & 02$\pm$0.05 & $-$1. & 772 & $-$1. & 776 & $-$1. & 80 & $-$1. & 71 & $-$2. & 02* & $-$2. & 06 \\ 
$\Delta^{++}$ & 3. & 7---7.5 & 5. & 389 & 5. & 448 & 5. & 23 & 4. & 51 & 6. & 04 & 4. & 52* \\ 
$\Delta^{+}$ &  & --- & 2. & 694 & 2. & 724 & 2. & 58 & 2. & 00 & 2. & 84 & 2. & 12 \\ 
$\Delta^{0}$ &  & --- & 0\phantom{.} &  & 0\phantom{.} &  & $-$0. & 078 & $-$0. & 51 & $-$0. & 36 & $-$0. & 29 \\ 
$\Delta^{-}$ &  & --- & $-$2. & 694 & $-$2. & 724 & $-$2. & 68 & $-$3. & 02 & $-$3. & 56 & $-$2. & 69 \\ 
$\Sigma^{*\ +}$ &  & --- & 2. & 927 & 3. & 040 & 3. & 05 & 2. & 69 & 3. & 07 & 2. & 63  \\ 
$\Sigma^{*\ 0}$ &  & --- & 0. & 275 & 0. & 316 & 0. & 289 & 0. & 02 & 0\phantom{.} &  & 0. & 08 \\ 
$\Sigma^{*\ -}$ &  & --- & $-$2. & 377 & $-$2. & 408 & $-$2. & 43 & $-$2. & 64 & $-$3. & 07 & $-$2. & 48 \\ 
$\Xi^{*\ 0}$ &  & --- & 0. & 536 & 0. & 632 & 0. & 68 & 0. & 54 & 0. & 36 & 0. & 44 \\ 
$\Xi^{*\ -}$ &  & --- & $-$2. & 067 & $-$2. & 092 & $-$2. & 13 & $-$1. & 87 & $-$2. & 56 & $-$2. & 27 \\ 
$\chi (\mu )$ &  & --- & 0. & 124 & 0. & 138 & 0. & 138 & 0. & 14 & 0. & 15 & 0. & 14 \\
\hline
\end{tabular}
\end{table}
%
\begin{table}[tbp]
\centering
\caption{Magnetic moments (in nuclear magnetons) obtained in other approaches as described in the text~-- continuation of Table~\ref{t3.4}.\label{t3.5}}
\begin{tabular}{cr@{}lr@{}lr@{}lr@{}lr@{}l}
\hline
Particles & \multicolumn{2}{c}{EXP~\cite{PDG12}} & \multicolumn{2}{c}{\cite{F09}} & \multicolumn{2}{c}{\cite{J11}} & \multicolumn{2}{c}{\cite{WL08,L98}} & \multicolumn{2}{c}{\cite{LWD91,LDW92}} \\ 
\hline
$P$ & 2. & 793 & 2. & 759 & 2. & 72 & 2. & 82$\pm$0.26 & 2. & 3$\pm$0.3 \\ 
$N$ & $-$1. & 913 & $-$1. & 975 & $-$1. & 91 & $-$1. & 97$\pm$0.15 & $-$1. & 3$\pm$0.2 \\ 
$\Lambda $ & $-$0. & 613$\pm$0.004 & $-$0. & 559 & $-$0. & 61 & $-$0. & 56$\pm$0.15 & $-$0. & 40$\pm$0.07 \\ 
$\Sigma^{+}$ & 2. & 458 & 2. & 428 & 2. & 45 & 2. & 31$\pm$0.25 & 1. & 9$\pm$0.2 \\ 
$\Sigma^{0}$ &  & --- & 0. & 625 & 0. & 64 & 0. & 69$\pm$0.07 & 0. & 54$\pm$0.09 \\ 
$\Sigma^{-}$ & $-$1. & 160$\pm$0.025 & $-$1. & 179 & $-$1. & 16 & $-$1. & 16$\pm$0.10 & $-$0. & 87$\pm$0.09 \\ 
$\Xi^{0}$ & $-$1. & 250$\pm$0.014 & $-$1. & 301 & $-$1. & 26 & $-$1. & 15$\pm$0.05 & $-$0. & 95$\pm$0.08 \\ 
$\Xi^{-}$ & $-$0. & 651$\pm$0.003 & $-$0. & 691 & $-$0. & 64 & $-$0. & 64$\pm$0.06 & $-$0. & 41$\pm$0.06 \\ 
$\left| \Sigma^{0}\rightarrow \Lambda \right| $ & 1. & 61$\pm$0.08 & 1. & 594 & 1. & 49 &  & --- & $-$1. & 15$\pm$0.16 \\ 
$\Omega^{-}$ & $-$2. & 02$\pm$0.05 & $-$2. & 042 & $-$2. & 03 & $-$1. & 49$\pm$0.45 & $-$1. & 40$\pm$0.10 \\ 
$\Delta^{++}$ & 3. & 7---7.5 & 5. & 390 & 5. & 64 & 4. & 13$\pm$1.30 & 4. & 91$\pm$0.61 \\ 
$\Delta^{+}$ &  & --- & 2. & 383 & 2. & 67 & 2. & 07$\pm$0.65 & 2. & 46$\pm$0.31 \\ 
$\Delta^{0}$ &  & --- & $-$0. & 625 & $-$0. & 30 & 0\phantom{.} &  & 0\phantom{.} &  \\ 
$\Delta^{-}$ &  & --- & $-$3. & 632 & $-$3. & 28 & $-$2. & 07$\pm$0.65 & $-$2. & 46$\pm$0.31 \\ 
$\Sigma^{*\ +}$ &  & --- & 2. & 519 & 2. & 97 & 2. & 13$\pm$0.82 & 2. & 05$\pm$0.26 \\ 
$\Sigma^{*\ 0}$ &  & --- & $-$0. & 303 & 0. & 05 & $-$0. & 32$\pm$0.15 & 0. & 27$\pm$0.05 \\ 
$\Sigma^{*\ -}$ &  & --- & $-$3. & 126 & $-$2. & 86 & $-$1. & 66$\pm$0.73 & $-$2. & 02$\pm$0.18 \\ 
$\Xi^{*\ 0}$ &  & --- & 0. & 149 & 0. & 41 & $-$0. & 69$\pm$0.29 & 0. & 46$\pm$0.07 \\ 
$\Xi^{*\ -}$ &  & --- & $-$2. & 596 & $-$2. & 45 & $-$1. & 51$\pm$0.52 & $-$1. & 68$\pm$0.12 \\ 
$\chi (\mu )$ &  & --- & 0. & 040 & 0. & 05 &  & --- &  & --- \\
\hline
\end{tabular}
\end{table}

When we pick the larger values of $Z_{0}$, the gain in the Casimir energy induces changes of bag parameters through the fitting procedure. For example, when the Casimir energy becomes larger, a smaller value of strong coupling constant is necessary to obtain the same masses of light hadrons chosen to fix the bag model parameters. Therefore we expect that with the increase of $Z_{0}$ the mass difference between vector and scalar mesons as well as between spin-3$/$2 and spin-1$/$2 baryons will become smaller. On the other hand, for the fit of the $\phi $ meson mass a larger strange quark mass will be necessary. Then the masses of hadrons containing strange quarks will increase. For some hadrons these two effects may partially compensate each other. We can see precisely such behaviour of hadron masses in Table~\ref{t3.2}: the masses of $K$, $\Xi $, and $\Omega^{-}$ increase, while the masses of $K^{*}$, $\Sigma^{*}$, and $\Xi^{*}$ fall down. An exception is the $\Sigma $ hyperon, the mass of which is strongly correlated with the mass of $\Lambda $ and the latter is nailed down because it (together with $\phi $) is used to obtain the model mass parameters. However, in the variant~\textit{1} of the model the changes of hadron masses are extremely small, and the values of $Z_{0}$ in a wide range ($0.1\leqslant Z_{0}\leqslant 0.6$) could be treated as a good choice (see Fig.~\ref{fig3.1}). We can say that in this sense the variant~\textit{1} is relatively stable. In the variant~\textit{2} the situation differs. An increase in $Z_{0}$ improves the light hadron mass spectrum (see Fig.~\ref{fig3.1} again). In this case the large values of $Z_{0}$ ($0.5\leqslant Z_{0}\leqslant 1$) would be preferable. Though we se that the reintroduction of the Casimir energy into the bag model Hamiltonian with the value of $Z_{0}\thickapprox 0.7$ can make the hadron mass predictions in the variant~\textit{2} of the model to be of the similar quality as the predictions of variant~\textit{1}.

Now let us see what is the effect of the changes in $Z_{0}$ on the calculated values of magnetic moments. From Figs.~\ref{fig3.3}, \ref{fig3.4} we see that there exists almost linear dependence of c.\,m.\,m. corrected magnetic moments on the Casimir energy parameter $Z_{0}$. For example, the choice $Z_{0} =$ 0.37 in variant~\textit{2} (see Fig.~\ref{fig3.3}) would provide the required value 2.79\,$\mu_{N}$ of the magnetic moment of a proton ($\mu_{N}$ stands for nuclear magneton, not to be confused with magnetic moment of neutron). The value 0.37 looks like magic because it is exactly the same value that was deduced from QED. No doubt, it is a simple coincidence. Such things happen.

Our strategy is to find the optimal values of the Casimir energy parameter from the minimization of $\chi (\mu )$. The values obtained in this way are $Z_{0}(\!\textit{Var1})=$ 0.22 and $Z_{0}(\!\textit{Var2})=$ 0.64 for the variants~\textit{1} and~\textit{2}, respectively. Now comparing predictions for magnetic moments in both variants of the model we see that, after all, variant~\textit{2} gives prominently better results. Explicit results for magnetic moments in both cases are listed in Table~\ref{t3.3}. And finally, in Tables~\ref{t3.4}, \ref{t3.5}
 we compare our predictions for magnetic moments with other calculations performed using different approaches. These are: the simple nonrelativistic result obtained using values from the two-parameter fit 
($\mu_{d}=-0.908 \mu_{N}$, $\mu_{s}=-0.592 \mu_{N}$, $\mu_{u}=-2 \mu_{d}$) 
 \cite{P96}, the translationally invariant model with harmonic oscillator wave functions \cite{PRS94,SPN95}, the chiral constituent quark model \cite{SDCG10}, the chiral perturbation theory \cite{GMAV08,GMV09}, the six-parameter fit based on the chiral bag sum rules \cite{H07}, the eight-parameter fit based on large-\textit{N}$_{\mathit{c}}$ chiral perturbation theory \cite{F09}, the eight-parameter fit based on $1/N_{c}$ expansion \cite{J11}, the QCD sum rules \cite{WL08,L98}, and the lattice calculations \cite{LWD91,LDW92}.

\section{ Discussion and conclusions}

In the end, let us see what we could expect and what we got. The main field of interest of our present work was the magnetic moments of the light baryons. The calculations were based on the usual quark model formula (Eq.~(\ref{eq2.05A})), which expresses the magnetic moments of baryons through the magnetic moments of individual quarks $\mu_{i}$. In the old-fashioned nonrelativistic approach $\mu_{i}$ usually appear as free parameters. If the isospin symmetry is assumed ($\bar{\mu}_{u}$ $=$ $\bar{\mu}_{d}$), one has a two-parameter fit. The typical deviation from the experiment $\chi (\mu )$ in such an approach \cite{P96,PRS94} is (0.13--0.14)\,$\mu_{N}$. In some models (for example, such as bag model, relativistic potential model \cite{BD83,BDD85,JR86,BJR90}, or QCD string approach \cite{KS00}) magnetic moments of quarks $\mu_{i}$ can be calculated directly without introduction of any new parameters. The agreement with experiment data in these cases is somewhat worse, with $\chi (\mu )$ in the interval (0.15--0.18)\,$\mu_{N}$. It is rather hard to improve this result. For example, in various chiral models \cite{SDCG10,GMAV08,H07} $\chi (\mu )$ lies in the range (0.14--0.15)\,$\mu_{N}$. Better fits require more free parameters. Some authors \cite{F09,J11} managed to reduce $\chi (\mu )$ to (0.04--0.05)\,$\mu_{N}$ with eight-parameter fits. It could be interesting to note that similar accuracy ($\chi (\mu )\thickapprox$ 0.05\,$\mu_{N}$) was achieved by L.G.~Pondrom in~\cite{P96} with only four free parameters in his hand-made expressions for magnetic moments of \textit{u-} and \textit{s-} quarks.

In our bag model calculations we have used only one free parameter (Casimir energy parameter $Z_{0}$) and obtained a significant improvement, though not so impressive as the 8-parameter fit. The discrepancy with experimental data $\chi (\mu )$ was reduced (in the variant~\textit{2} of the model) from 0.21\,$\mu_{N}$ to 0.12\,$\mu_{N}$. Simultaneously, predictions for the light hadron mass spectrum were improved as well. Moreover, the value of the Casimir energy parameter $Z_{0}$ required for the ``best'' fit is close to its theoretical prediction \cite{M83}.

So, should we include the Casimir energy in the bag model Hamiltonian or not? From  theoretical point of view the answer seems to be ``yes''. From phenomenological point of view the answer would be more modest: ``it depends''. If one is interested in the mass spectrum only, one can use the simpler variant~\textit{1} of the bag model and plausibly do not care about anything else. However, if for some reason we have decided to exploit the theoretically more consistent variant~\textit{2}, or we are interested in the calculation of magnetic moments, then incorporation of the Casimir energy in the bag model Hamiltonian would be a reasonable choice.


\begin{thebibliography}{99}
\bibitem{LT11}
R.F.~Lebed and R.H.~TerBeek, Phys. Rev.~D \textbf{83}, 016009 (2011), \\
\href{http://dx.doi.org/10.1103/PhysRevD.83.016009}{http://dx.doi.org/10.1103/PhysRevD.83.016009}

\bibitem{DJJK75}
T.~DeGrand, R.L.~Jaffe, K.~Johnson, and J.~Kiskis, Phys. Rev.~D \textbf{12}, 2060 (1975), \\
\href{http://dx.doi.org/10.1103/PhysRevD.12.2060}{http://dx.doi.org/10.1103/PhysRevD.12.2060}

\bibitem{HK78}
P.~Hasenfratz and J.~Kuti, Phys. Rep. \textbf{40}, 75 (1978), \\
\href{http://dx.doi.org/10.1016/0370-1573(78)90076-5}{http://dx.doi.org/10.1016/0370-1573(78)90076-5}

\bibitem{MK01}
P.~Mi\v{s}kinis and G.~Karlikauskas, Nucl. Phys.~A \textbf{683}, 339 (2001), \\
\href{http://dx.doi.org/10.1016/S0375-9474(00)00442-5}{http://dx.doi.org/10.1016/S0375-9474(00)00442-5}

\bibitem{HDDT78}
R.H.~Hackman, N.G.~Deshpande, D.A.~Dicus, and V.L.~Teplitz, Phys. Rev.~D \textbf{18}, 2537 (1978), \\
\href{http://dx.doi.org/10.1103/PhysRevD.18.2537}{http://dx.doi.org/10.1103/PhysRevD.18.2537}

\bibitem{HK82}
A.~Halprin and A.K.~Kerman, Phys. Rev.~D \textbf{26}, 2532 (1982), \\
\href{http://dx.doi.org/10.1103/PhysRevD.26.2532}{http://dx.doi.org/10.1103/PhysRevD.26.2532}

\bibitem{BSRMT84}
J.~Bartelski, A.~Szymacha, Z.~Ryzak, L.~Mankiewicz, and S.~Tatur, Nucl. Phys.~A \textbf{424}, 484 (1984), \\
\href{http://dx.doi.org/10.1016/0375-9474(84)90006-X}{http://dx.doi.org/10.1016/0375-9474(84)90006-X}

\bibitem{S98}
V.~\v{S}imonis, Lith. J.~Phys. \textbf{38}, 274 (1998)

\bibitem{BS04}
A.~Bernotas and V.~\v{S}imonis, Nucl. Phys.~A \textbf{741}, 179 (2004), \\
\href{http://dx.doi.org/10.1016/j.nuclphysa.2004.05.017}{http://dx.doi.org/10.1016/j.nuclphysa.2004.05.017}

\bibitem{CHP83}
C.E.~Carlson, T.H.~Hansson, and C.~Peterson, Phys. Rev.~D \textbf{27}, 1556 (1983), \\
\href{http://dx.doi.org/10.1103/PhysRevD.27.1556}{http://dx.doi.org/10.1103/PhysRevD.27.1556}

\bibitem{M83}
K.A.~Milton, Phys. Rev.~D \textbf{27}, 439 (1983), \\
\href{http://dx.doi.org/10.1103/PhysRevD.27.1556}{http://dx.doi.org/10.1103/PhysRevD.27.1556}

\bibitem{CJJTW74}
A.~Chodos, R.L.~Jaffe, K.~Johnson, C.B.~Thorn, and V.F.~Weisskopf, Phys. Rev.~D \textbf{9}, 3471 (1974), \\
\href{http://dx.doi.org/10.1103/PhysRevD.9.3471}{http://dx.doi.org/10.1103/PhysRevD.9.3471}

\bibitem{DJ80}
J.F.~Donoghue and K.~Johnson, Phys. Rev.~D \textbf{21}, 1975 (1980), \\
\href{http://dx.doi.org/10.1103/PhysRevD.21.1975}{http://dx.doi.org/10.1103/PhysRevD.21.1975}

\bibitem{BSMT84}
J.~Bartelski, A.~Szymacha, L.~Mankiewicz, and S.~Tatur, Phys. Rev.~D \textbf{29}, 1035 (1984), \\
\href{http://dx.doi.org/10.1103/PhysRevD.29.1035}{http://dx.doi.org/10.1103/PhysRevD.29.1035}

\bibitem{TT85}
D.~Tadi\'{c}, G.~Tadi\'{c}, Phys. Rev.~D \textbf{31}, 1700 (1985), \\
\href{http://dx.doi.org/10.1103/PhysRevD.31.1700}{http://dx.doi.org/10.1103/PhysRevD.31.1700}

\bibitem{HM90}
L.C.L.~Hollenberg and B.H.J.~McKellar, J.~Phys.~G \textbf{16}, 31 (1990), \\
\href{http://dx.doi.org/10.1088/0954-3899/16/1/006}{http://dx.doi.org/10.1088/0954-3899/16/1/006}

\bibitem{M01}
K.A.~Milton, \emph{The Casimir Effect} (World Scientific, Singapore, 2001), \\
\href{http://www.worldscibooks.com/physics/4505.html}{http://www.worldscibooks.com/physics/4505.html}

\bibitem{LR96}
S.~Leseduarte and A.~Romeo, Ann. Phys. (NY) \textbf{250}, 448 (1996), \\
\href{http://dx.doi.org/10.1006/aphy.1996.0101}{http://dx.doi.org/10.1006/aphy.1996.0101}

\bibitem{PDG12}
K.~Nakamura et al. (Particle Data Group), J.~Phys.~G \textbf{37}, 075021 (2010), \\
\href{http://dx.doi.org/10.1088/0954-3899/37/7A/075021}{http://dx.doi.org/10.1088/0954-3899/37/7A/075021}

\bibitem{P96}
L.G.~Pondrom, Phys. Rev.~D \textbf{53}, 5322 (1996), \\
\href{http://dx.doi.org/10.1103/PhysRevD.53.5322}{http://dx.doi.org/10.1103/PhysRevD.53.5322}

\bibitem{PRS94}
A.R.~Panda, K.C.~Roy, and R.K.~Sahoo, Phys. Rev.~D \textbf{49}, 4659 (1994), \\
\href{http://dx.doi.org/10.1103/PhysRevD.49.4659}{http://dx.doi.org/10.1103/PhysRevD.49.4659}

\bibitem{SPN95}
R.K.~Sahoo, A.R.~Panda, and A.~Nath, Phys. Rev.~D \textbf{52}, 4099 (1995), \\
\href{http://dx.doi.org/10.1103/PhysRevD.52.4099}{http://dx.doi.org/10.1103/PhysRevD.52.4099}

\bibitem{SDCG10}
N.~Sharma, H.~Dahiya, P.K.~Chatley, and M.~Gupta, Phys. Rev.~D \textbf{81}, 073001 (2010), \\
\href{http://dx.doi.org/10.1103/PhysRevD.81.073001}{http://dx.doi.org/10.1103/PhysRevD.81.073001}

\bibitem{GMAV08}
L.S.~Geng, J.~Martin Camalich, L.~Alvarez-Ruso, and M.J.~Vicente Vacas, Phys. Rev. Lett. \textbf{101}, 222002 (2008), \\
\href{http://dx.doi.org/10.1103/PhysRevLett.101.222002}{http://dx.doi.org/10.1103/PhysRevLett.101.222002}

\bibitem{GMV09}
L.S.~Geng, J.~Martin Camalich, and M.J.~Vicente Vacas, Phys. Rev.~D \textbf{80}, 034027 (2009), \\
\href{http://dx.doi.org/10.1103/PhysRevD.80.034027}{http://dx.doi.org/10.1103/PhysRevD.80.034027}

\bibitem{H07}
S-T.~Hong, Phys. Rev.~D \textbf{76}, 094029 (2007), \\
\href{http://dx.doi.org/10.1103/PhysRevD.76.094029}{http://dx.doi.org/10.1103/PhysRevD.76.094029}

\bibitem{F09}
R.~Flores-Mendieta, Phys. Rev.~D \textbf{80}, 094014 (2009), \\
\href{http://dx.doi.org/10.1103/PhysRevD.80.094014}{http://dx.doi.org/10.1103/PhysRevD.80.094014}

\bibitem{J11}
E.~Jenkins, arXiv:1111.2055v2 (2011), \\
\href{http://lanl.arxiv.org/abs/1111.2055v2}{http://lanl.arxiv.org/abs/1111.2055v2}

\bibitem{WL08}
L.~Wang and F.X.~Lee, Phys. Rev.~D \textbf{78}, 013003 (2008), \\
\href{http://dx.doi.org/10.1103/PhysRevD.78.013003}{http://dx.doi.org/10.1103/PhysRevD.78.013003}

\bibitem{L98}
F.X.~Lee, Phys. Rev.~D \textbf{57}, 1801 (1998), \\
\href{http://dx.doi.org/10.1103/PhysRevD.57.1801}{http://dx.doi.org/10.1103/PhysRevD.57.1801}

\bibitem{LWD91}
D.B.~Leinweber, R.M.~Woloshyn, and T.~Draper, Phys. Rev.~D \textbf{43}, 1659 (1991), \\
\href{http://dx.doi.org/10.1103/PhysRevD.43.1659}{http://dx.doi.org/10.1103/PhysRevD.43.1659}

\bibitem{LDW92}
D.B.~Leinweber, T.~Draper, and R.M.~Woloshyn, Phys. Rev.~D \textbf{46}, 3067 (1992), \\
\href{http://dx.doi.org/10.1103/PhysRevD.46.3067}{http://dx.doi.org/10.1103/PhysRevD.46.3067}

\bibitem{BD83}
N.~Barik and M.~Das, Phys. Rev.~D \textbf{28}, 2823 (1983), \\
\href{http://dx.doi.org/10.1103/PhysRevD.28.2823}{http://dx.doi.org/10.1103/PhysRevD.28.2823}

\bibitem{BDD85}
N.~Barik, B.K.~Dash, and M.~Das, Phys. Rev.~D \textbf{31}, 1652 (1985), \\
\href{http://dx.doi.org/10.1103/PhysRevD.31.1652}{http://dx.doi.org/10.1103/PhysRevD.31.1652}

\bibitem{JR86}
S.N.~Jena and D.P.~Rath, Phys. Rev.~D \textbf{34}, 196 (1986), \\
\href{http://dx.doi.org/10.1103/PhysRevD.34.196}{http://dx.doi.org/10.1103/PhysRevD.34.196}

\bibitem{BJR90}
N.~Barik, S.N.~Jena, and D.P.~Rath, Phys. Rev.~D \textbf{41}, 1568 (1990), \\
\href{http://dx.doi.org/10.1103/PhysRevD.41.1568}{http://dx.doi.org/10.1103/PhysRevD.41.1568}

\bibitem{KS00}
B.O.~Kerbikov and Yu.A.~Simonov, Phys. Rev.~D \textbf{62}, 093016 (2000), \\
\href{http://dx.doi.org/10.1103/PhysRevD.62.093016}{http://dx.doi.org/10.1103/PhysRevD.62.093016}

\end{thebibliography}
\end{document}